\documentclass{sig-alternate}

\usepackage{times}
\usepackage{booktabs, multicol, multirow}
\usepackage{graphicx}
\usepackage{amsmath}
\usepackage{algorithm2e}
\usepackage{subfigure}
\usepackage{mdwlist}
\usepackage{color}
\usepackage{bbold}
\usepackage{url}
\usepackage{tikz}
\usepackage{xkeyval}
\usetikzlibrary{shapes,snakes}
\usepackage{boiboites}
\usepackage{etoolbox}
\usepackage{flushend}

\renewcommand{\vec}[1]{\mathbf{#1}}

\global\long\def\x{{\vec{x}}}

\global\long\def\matR{\mathbf{R}}
\global\long\def\matQ{\mathbf{Q}}

\global\long\def\matF{\mathbf{F}}

\newcommand{\methodLong}{\textsc{MovieDesign}\xspace}
\newcommand{\method}{\textsc{MD}\xspace}
\newcommand{\featmodelTitle}{{\LARGE{\textsc{PNP}}}\xspace}
\newcommand{\randomWalk}{\textsc{PNP}\xspace}
\newcommand{\featmodel}{\textsc{PNP}\xspace}

\newcommand{\featmodelT}{{\large\textsc{PNP}}\xspace}
\newcommand{\positive}{\textsc{Positive}\xspace}
\newcommand{\negative}{\textsc{Negative}\xspace}

\newcommand{\onevector}{\mathbb{1}}

\newcommand{\popular}{\textsc{Popular}\xspace}
\newcommand{\topF}{\textsc{Top}\xspace}
\newcommand{\knn}{k\textsc{NN}\xspace}
\newcommand{\wknn}{\textsc{w-}k\textsc{NN}\xspace}

\newcommand{\userSet}{\mathcal{U}}
\newcommand{\movieSet}{\mathcal{M}}
\newcommand{\featureSet}{\mathcal{F}}
\newcommand{\selectedSet}{\mathcal{S}}
\newcommand{\rMat}{\mathbf{R}}
\newcommand{\fMat}{\mathbf{F}}
\newcommand{\wMat}{\mathbf{W}}

\newcommand{\m}{m}
\newcommand{\users}{u}
\newcommand{\f}{f}
\newcommand{\thres}{\tau}
\newcommand{\wvec}{ {\vec{w} }}
\newcommand{\xvec}{ {\vec{x} }}

\newcommand{\hide}[1]{}
\newcommand{\bit}{\begin{itemize}}
\newcommand{\eit}{\end{itemize}}

\newcommand{\new}[1]{ {\color{blue} {#1} } }
\renewcommand{\new}[1]{ {#1} }

     \newcounter{intuition}
     
     \newsavebox{\coloredbgbox}

     \newcounter{observation}
     
     \newenvironment{observation}{
     	\begin{lrbox}{\coloredbgbox}
     		\begin{minipage}{\dimexpr1\linewidth-2\fboxsep-2\fboxrule\relax}
     			\refstepcounter{observation}
     			\textbf{\textsc{Observation} \theobservation.}
     		}
       		{
     		\end{minipage}
     	\end{lrbox}
     	\begin{center}
     		\fcolorbox{black!30}{black!02}{\usebox{\coloredbgbox}}
     	\end{center}
     }

   \newcounter{problem}
       
       \newenvironment{problem}{
       	\begin{lrbox}{\coloredbgbox}
       		\begin{minipage}{\dimexpr1\linewidth-2\fboxsep-2\fboxrule\relax}
       			\refstepcounter{problem}
       			\textbf{\textsc{Problem} \theproblem.}
       		}
         		{
       		\end{minipage}
       	\end{lrbox}
       	\begin{center}
       		\fcolorbox{black}{black!06}{\usebox{\coloredbgbox}}
       	\end{center}
       }

          \newcounter{definition}
              
              \newenvironment{definition}{
              	\begin{lrbox}{\coloredbgbox}
              		\begin{minipage}{\dimexpr1\linewidth-2\fboxsep-2\fboxrule\relax}
              			\refstepcounter{definition}
              			\textbf{\textsc{Definition}.}
              		}
                		{
              		\end{minipage}
              	\end{lrbox}
              	\begin{center}
              		\fcolorbox{black!25}{black!03}{\usebox{\coloredbgbox}}
              	\end{center}
              }

\begin{document}

\title{\featmodelTitle: Fast Path Ensemble Method for  Movie Design}

\newfont{\smfnt}{phvr8t at 11pt}

\def\alignauthorL{
\end{tabular}
  \begin{tabular}[t]{p{1.21\auwidth}}\centering}

\numberofauthors{6} 
\author{
\alignauthor
 Danai Koutra\titlenote{This work started during an internship at Technicolor.} \\
 \affaddr{University of Michigan}\\
\email{\{dkoutra\}@umich.edu}
\alignauthor
Abhilash Dighe \\
\affaddr{University of Michigan}\\
 \email{ adighe@umich.edu}
\alignauthorL
Smriti Bhagat$^\dagger$, Udi Weinsberg$^\dagger$\\
\affaddr{Facebook}\\
 \email{\{smr, udi\}@fb.com }
\and
\alignauthor 
Stratis Ioannidis\titlenote{This work started while at Technicolor.}\\
       \affaddr{Northeastern University}\\
       \email{ioannidis@neu.edu}
\alignauthor 
Christos Faloutsos\\
       \affaddr{Carnegie Mellon University}\\
       \email{christos@cs.cmu.edu}
\alignauthorL 
Jean Bolot\\
       \affaddr{Technicolor}\\
       \email{  jean.bolot@technicolor.com}
}

\maketitle

\begin{abstract}
How can we design a product or movie that will attract, for example, the interest of Pennsylvania adolescents or liberal newspaper critics? What should be the genre of that movie and who should be in the cast? 
In this work, we seek to identify how we can design \emph{new} movies with features tailored to a specific user population.
We formulate the movie design as an optimization problem over the inference of user-feature scores and selection of the features that maximize the number of attracted users. 
Our approach, \featmodel, is based on a heterogeneous, tripartite graph of users, movies and features (e.g., actors, directors, genres), where users rate movies and features contribute to movies. We learn the preferences by leveraging user similarities defined through different types of relations, 
 and show that our method outperforms state-of-the-art approaches, including matrix factorization and other heterogeneous graph-based analysis. 
We evaluate \featmodel on publicly available real-world data 
and show that it is highly scalable and effectively provides movie designs oriented towards different groups of users, including men, women, and adolescents.

\end{abstract}

\section{Introduction}
\label{sec:intro}

Creating products that satisfy the market is critical to companies 
as it determines their success and revenue~\cite{Andreessen07}. Online services, such as Amazon, eBay, and Netflix,   use data-driven approaches to recommend products to their customers. Recently, however, companies like Netflix have employed their direct knowledge of user viewing habits to make licensing and original programming decisions; this was indeed the case with, e.g., `House of Cards', one of their most-watched shows~\cite{carr2013giving,algoprog}.
Can we take this decision-making process one step further, and leverage the large amounts of available data (e.g., user ratings or (dis)likes, reviews,  product characteristics) \textit{to inform the design of new products and services} that will likely satisfy the target customers? 

\new{Market fit for products and services~\cite{Andreessen07}, team formation~\cite{LappasLT09,AnagnostopoulosBCGL12}, 
and team replacement~\cite{LiTCELB15} are related problems with a similar  goal: \textit{designing} a new product, service, or team that will succeed, given a target market or task.  The large number of  possible choices, combinations, and constraints, as well as the challenge of assessing audience demand, make this a considerably hard problem. Currently, experts use their judgement to estimate solutions to this problem, but this does not scale or allow leveraging massive datasets.}

  \begin{figure}[t!]
  \vspace{-0.1cm}
       \centering
       \includegraphics[width=0.75\columnwidth]{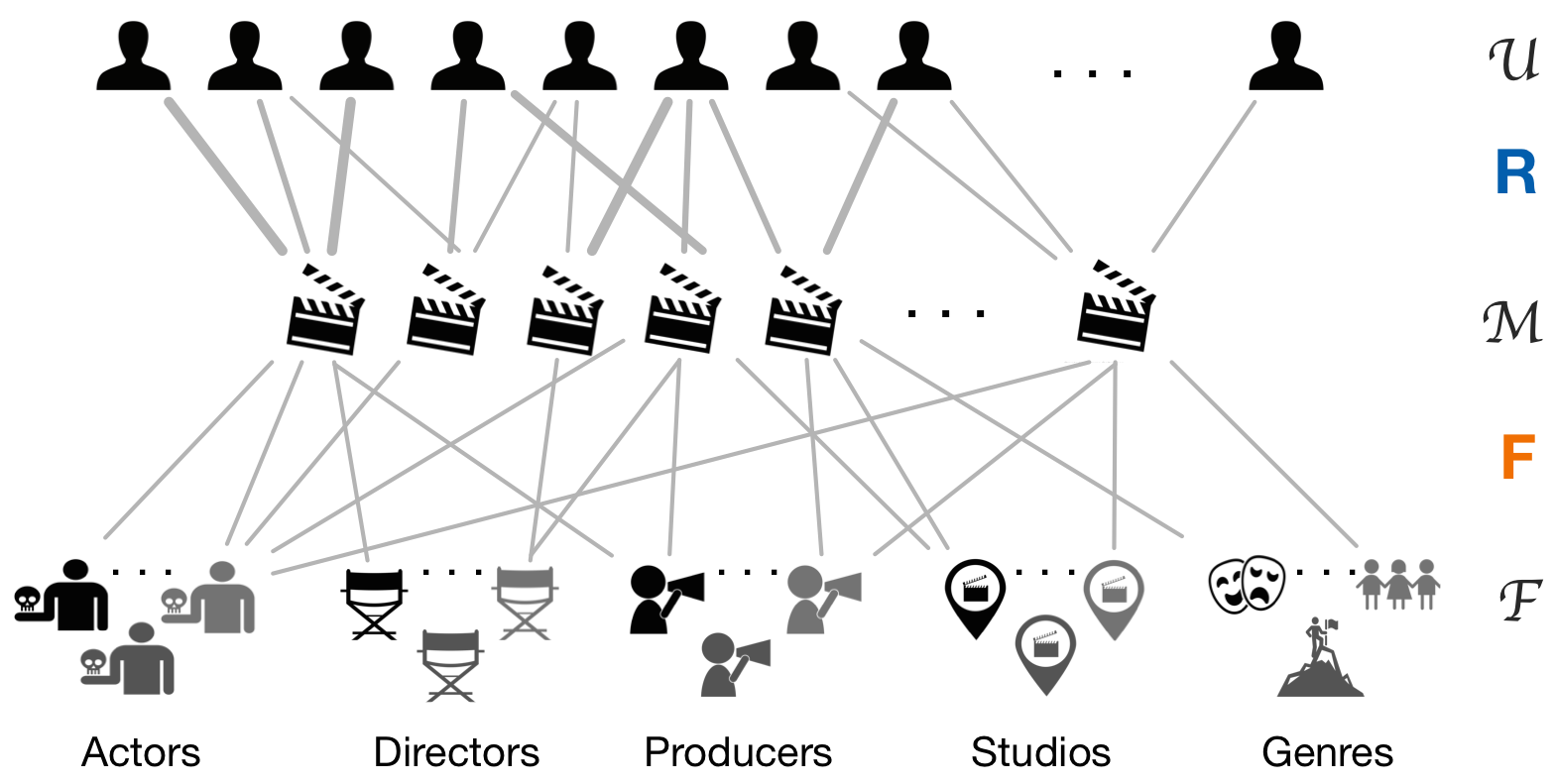}
       \label{fig:tripartite}
     \vspace{-0.3cm}
  	 \caption{Tripartite graph of the target users $\mathcal{U}$, their rated movies $\mathcal{M}$ 
  	 and the movie features $\mathcal{F}$.} 
     \vspace{-0.2cm}
   \end{figure}

 The goal of this paper is to formalize the movie design problem and solve it in a tractable and scalable fashion. 
    Our motivation comes from the movie entertainment industry: 
    We investigate how to design a successful movie that 
will attract the interest of an arbitrary targeted demographic. 
 Specifically, we tackle the following problem (in short, \method):

\vspace{-0.2cm}  
  \begin{problem}[\methodLong or \method{}] 
  \label{prob:featureDependent}
   \textbf{Given} (a) a set of users, target users, movies, and features, 
   (b) some user movie ratings, and 
   (c) the feature-movie memberships; 
 \textbf{design}  a movie which will attract most of the target users. 
  \end{problem}
  \vspace{-0.2cm}

Given additional data about the cost of each feature, the problem can be augmented to handle budget constraints as well. 
The \method problem resembles 
recommender systems~\cite{MelvilleMN02,LindenS03,HuKD08,YuanCZ11,Ronen2013RECSYS} and group recommendations \cite{OConnorCKR01,JamesonS07}. Its main difference is that it does not aim to find the best \textit{existing} movie that the target users are likely to enjoy. Instead, given the user preferences, it determines the features for a \textit{new} movie that is likely to attract the most users. 

Solving \method poses several challenges.  Clearly, 
identifying an effective design relies on how movie features affect user preferences. As extensively documented in the recommendation literature \cite{koren2009matrix,bennett2007netflix}, and also corroborated by our experiments, ``collaborative filtering'' approaches, exploiting similarities between users, tend to significantly outperform regressing individual user preferences in isolation. It is not immediately clear how to solve \method by leveraging both explicit features and user similarities  in a principled, tractable fashion. 
An additional challenge arises from the fact that the \method problem can be applied to an arbitrary set of users $\mathcal{U}$. Indeed, different strategists may wish to attract varied demographics, and our design should be able to cope with arbitrary, repeated requests. As a result, we would like to  efficiently determine movie designs  targeting \textit{any} given group.

To address these challenges,  we propose a graph-theoretic approach for the \method problem 
and contribute:

\noindent \textbf{$\bullet$ Problem Formulation}: We formally pose \method as a problem for designing successful new movies for a target audience. Moreover, we adopt an approach that separates the problem solution into two phases: user-feature preference inference, and model-based design. 
This separation allows us to infer user preferences and efficiently handle arbitrary ``design queries'', as defined by targeted user groups.

\noindent \textbf{$\bullet$ Path-Based Training}:  To infer the user-feature scores, we propose a novel model  based on predefined walks on a heterogeneous graph consisting of users, movies, and features, which treats ``dislikes'' in a natural way. This methodology allows us to leverage both features \emph{and} user similarity. 
Although we focus on the setting of movies, our method is generalizable  to any setting where user ratings and product features are available.

\noindent \textbf{$\bullet$ Model-Based Design}: Having inferred user-feature preferences, we formulate the selection of features that compose a movie as an optimization problem with capacity and other constraints, and establish conditions under which the problem solution is tractable.

\noindent \textbf{$\bullet$ Experiments:} By using real-world data with $\sim$5 \textit{million} movie-ratings and 175\,000 movie-features, we show that our model of user behavior succeeds in describing user preferences with very high accuracy. In  addition, combined with our optimization method, it results in movie design choices that significantly outperform those of the competitors. 

The paper is organized as follows: 
We describe our proposed method, \featmodel, the dataset, and the experimental evaluation in Sections~\ref{sec:featuremodel}, ~\ref{sec:data}, and~\ref{sec:experiments}, respectively. 
Sections~\ref{sec:related} and~\ref{sec:conclusions} present the related work and our conclusions.

\section{Proposed Method: \featmodelT}
\label{sec:featuremodel}
For the \method problem, we assume that the data consists of: a set of users $\mathcal{U}$, 
 a set of movies $\mathcal{M}$, 
 and a set of movie features $\mathcal{F}$. 
 In addition, the feature set $\mathcal{F}$ is partitioned into types (e.g., actors, directors, genres): we denote by $\mathcal{T}$ 
  the set of types, and by $\mathcal{F}_\ell\subset\mathcal{F}$ the set of features of type $\ell\in \mathcal{T}$. 
The relations between entities consist of: (a) the user-movie ratings, containing  tuples of the form $(i,j,r_{ij})$, where $i\in \mathcal{U}$ and $j\in \mathcal{M}$, 
and are organized in a  $u \times m$ matrix $\mathbf{R}$, with zeros indicating absent ratings; 
(b) the movie-feature memberships stored in a $m \times f$ binary matrix $\mathbf{F}$, where a non-zero entry $(j,k)$ means that feature $k$ belongs to movie $j$. Throughout the paper, following observations in the literature~\cite{DernbachTKWDA16}, we say that user $i$ `likes' movie $j$ 
if the rating $r_{ij}$ is larger than $i$'s average  rating $\bar{r}_{i}$ among non-zero entries in $\mathbf{R}$. 
Table~\ref{tab:definitions}
presents the most commonly used symbols.

The goal of the \method problem (Problem~\ref{prob:featureDependent}) is to design a new movie by selecting its features so that it is 
 liked by as many of the people in the target audience as possible. To make the problem tractable, and handle the challenges that we described in the introduction, we reduce it to solving  two separate subproblems. We tackle these subproblems in Sections~\ref{sec:step1}, and~\ref{seq:step2}, respectively.

\vspace{-0.2cm}
  \begin{problem}[Inferring User-Feature Scores] 
  \label{prob:step1}
     \textbf{Given} (a) a set of users $\mathcal{U}$, movies $\mathcal{M}$, and features $\mathcal{F}$,
     (b) the user movie ratings $\matR$, and 
     (c) the movie-feature memberships $\mathbf{F}$;
     \textbf{find} the user-feature preference scores. 
  \end{problem}
  
\vspace{-0.2cm}
The first subproblem amounts to learning user-feature preference scores, capturing the propensity of a given user to like a movie that contains a given feature; the higher the score, the more likely a user is to like this movie. In the second subproblem, we use these scores to formulate the objective of maximizing the number of users liking the movie, among a targeted set.  
\vspace{-0.2cm}

    \begin{problem}[Designing the Movie] 
    \label{prob:step2}
    \textbf{Given} (a) a set of target users $\mathcal{U'}$, 
                               features $\mathcal{F}$, and types $\mathcal{T}$, and
         (b) the user-feature preferences; 
     \textbf{select}  features for a new movie  s.t. 
                      the number of users who will probably like it is maximized.
    \end{problem}

 \vspace{-0.2cm}
If additionally each feature $k$ is associated with a cost $c_k$, we can augment the problem by incorporating budget constraints. 
These constraints would express the requirement that the total expense on each feature type (actor, director, etc.) does not exceed  a budget $B_\ell$ per type.
Next we discuss our solutions to these two problems.

\begin{table}[t!]
\vspace{-0.4cm}
\centering
\caption{Symbols and Definitions. Bold capital: matrices; Lowercase: vectors; Plain font: scalars.}
\label{tab:definitions}
\resizebox{\columnwidth}{!}{
\begin{tabular}{c l} 
\toprule
\multicolumn{1}{l}{\textbf{Symbol}} & \multicolumn{1}{l}{\textbf{Description}} \\ 
\midrule
\multicolumn{1}{l}{$G$} & tripartite input graph \\ 
\multicolumn{1}{l}{$\userSet (\userSet'), \movieSet,\featureSet, \mathcal{T}$} & the set of (target) users, movies, features, and types, resp.  \\ 
\multicolumn{1}{l}{$u (u'), m, f, t$} & number of (target) users, movies, features, and types, resp.  \\ \hline
\multicolumn{1}{l}{$\rMat$} & $\users \times \m$ matrix of user-movie ratings, with elements $r_{ij}$ \\ 
\multicolumn{1}{l}{$\fMat$} & $\m \times \f$ movie-feature membership matrix, with elements $f_{ij}$ \\ 
\multicolumn{1}{l}{$\mathbf{W}$} & inferred $\users \times \f$ matrix of user-feature preferences \\ 
\multicolumn{1}{l}{$\mathbf{Q}$} & proposed $\users \times \m$ matrix of user-movie ratings, with elmnts. $q_{ij}$ \\  \hline 
\multicolumn{1}{l}{$\thres$} & threshold in the linear threshold model \\ 
\multicolumn{1}{l}{$\selectedSet$} & the set  of chosen features  \\ 
\multicolumn{1}{l}{$G(S)$} & user conversion function  \\ 
\multicolumn{1}{l}{$\xvec$} & $1 \times \f$ binary characteristic vector of $\mathcal{S}$  \\
\multicolumn{1}{l}{$c_k$} & cost per feature $k$ (e.g., salary)  \\
\multicolumn{1}{l}{$B_\ell$} & budget per type $\ell$  \\
\bottomrule 
\end{tabular}
}
\vspace{-0.5cm}
\end{table}

\begin{figure*}[t!]
\centering
\includegraphics[width=0.98\textwidth]{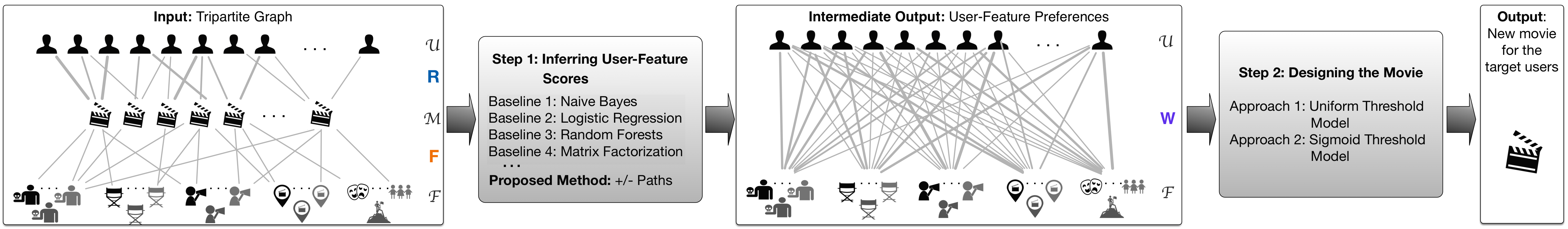}
\vspace{-0.2cm}
\caption{\method: Proposed pipeline for designing a new movie based on user ratings and movie features.}\label{fig:input_pipeline_output}
\label{fig:pipeline}
\end{figure*}

\begin{figure*}[t]
        \centering
        \subfigure[Direct 2-step Path]{\label{fig:twostep} \includegraphics[width=0.3\textwidth]{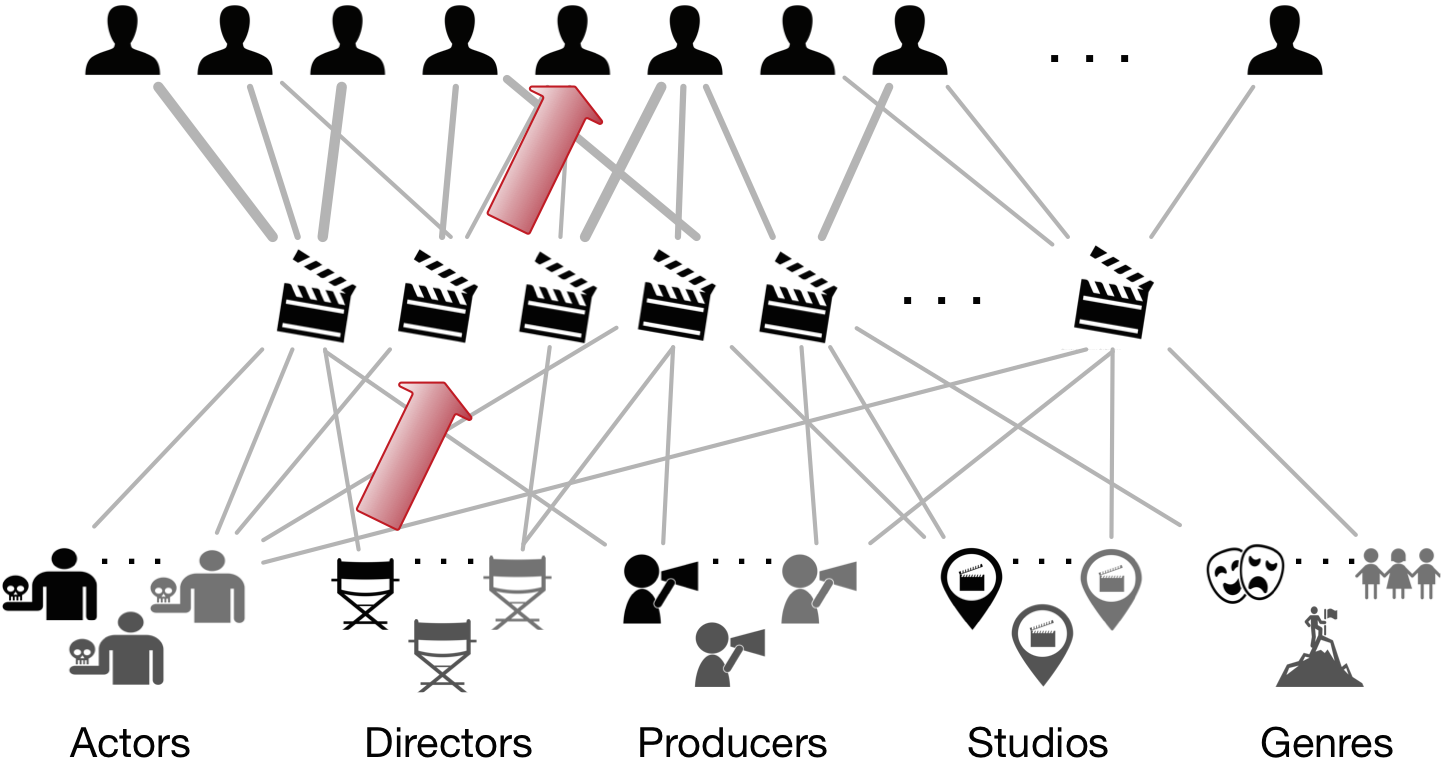}}
        ~~~~~
        \subfigure[User-based 4-step Path]{\label{fig:fourstep1} \includegraphics[width=0.3\textwidth]{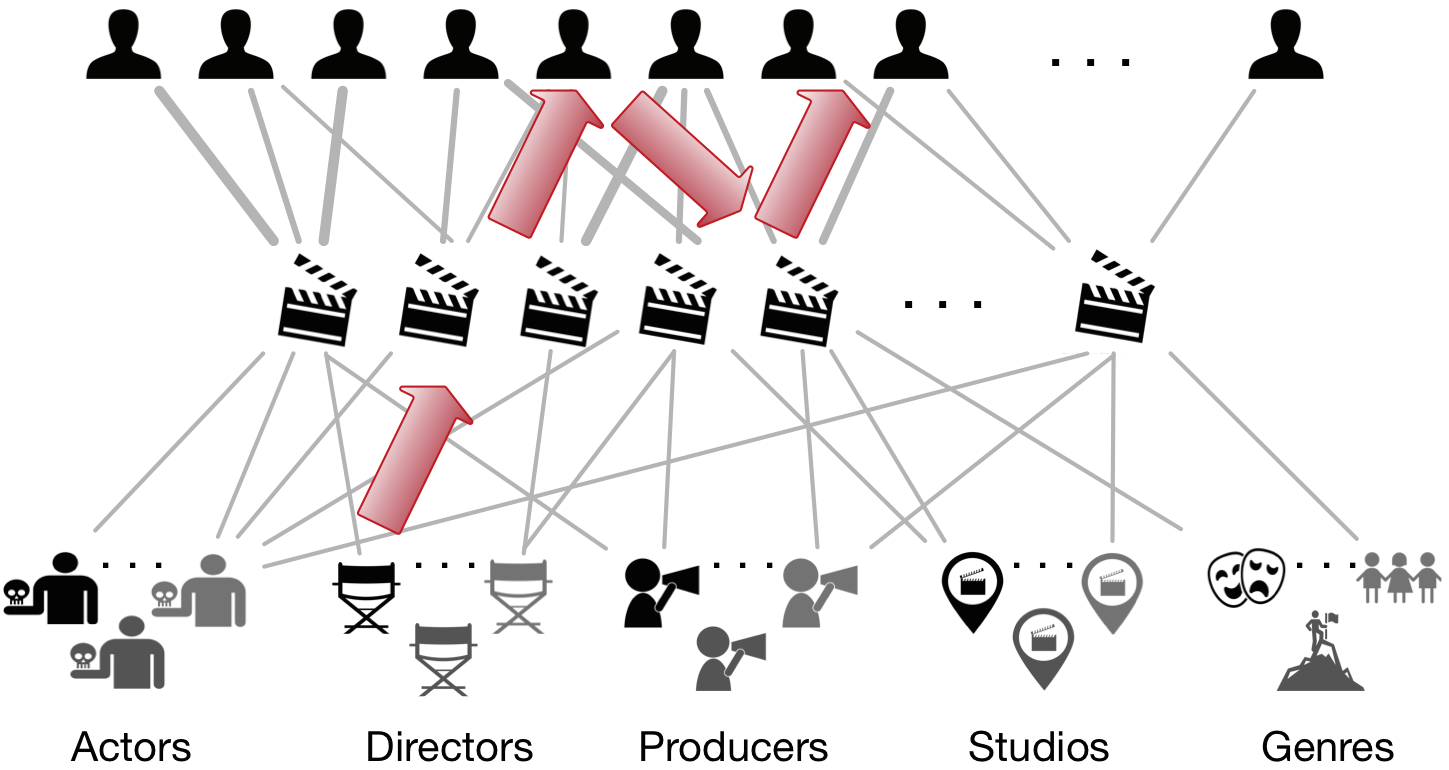}}
        ~~~~~
        \subfigure[Feature-based 4-step Path]{\label{fig:fourstep2} \includegraphics[width=0.3\textwidth]{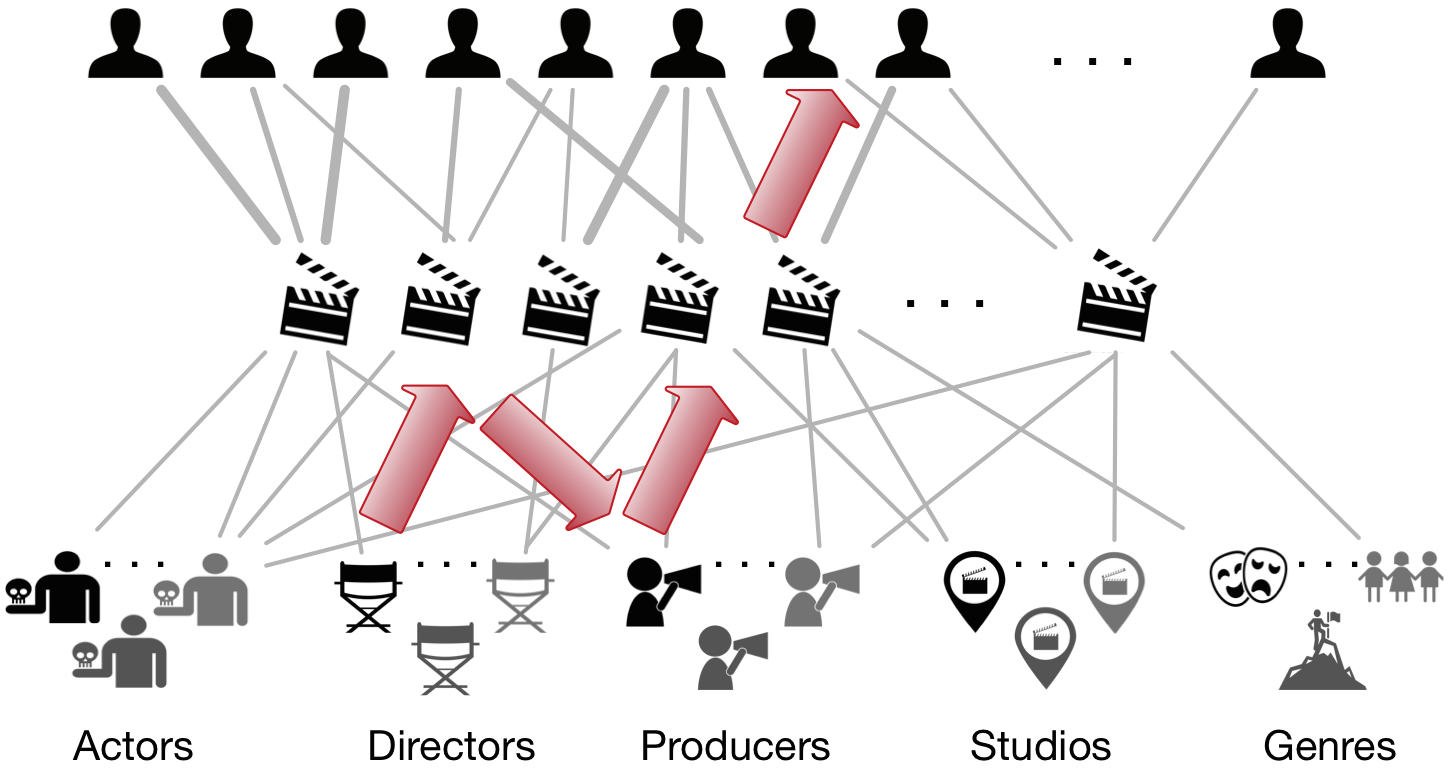}}
        \vspace{-0.2cm}
        \caption{Combination of three types of predefined walks on the heterogeneous tri-partite graph. The red arrows show the predefined direction of each walk.}\label{fig:pathtypes}
        \vspace{-0.2cm}
\end{figure*}

\subsection{Step 1. Inferring user-feature scores}
\label{sec:step1}

The first constituent problem of \method, Problem~\ref{prob:step1}, can be approached as a classification problem, in which binary `like' labels of every user are regressed from movie features. Traditional methods (e.g., random forests, matrix factorization) can be used to solve such a problem; as a byproduct, these methods often quantify the effect of each individual feature on a user's decision. However, these methods do not perform well (cf. Section~\ref{sec:experiments}) as they either ignore commonalities between users or make use of latent features which cannot be used to identify user-feature preferences.

We propose to solve the \method problem by handling both constituent problems via a graph-theoretic approach, which cleverly leverages \emph{both} explicit features \emph{and} user similarities, and, most importantly, its resulting scores also lead to a tractable movie design optimization for an arbitrary set of target users (Section~\ref{seq:step2}). We model the input data as an undirected, heterogeneous, tri-partite network 
(Figure~\ref{fig:tripartite}), where the nodes are users, movies, and movie features. The relationships or edge types are `rated-by', 
 and `belong-to'. We give a pictorial overview of our approach in Figure~\ref{fig:pipeline}.

\enlargethispage{\baselineskip}
Specifically, our proposed method, \featmodel (Positive / Negative Paths), infers the user-feature scores by performing walks of fixed length on the 
tripartite graph of users, movies and features, thus leveraging information about user-movie ratings as in `collaborative' approaches, as well as movie `content' (features). 
\featmodel  is based on meta-paths, which was first introduced for similarity search in heterogeneous information networks~\cite{SunHYYW11}. Informally, a meta-path consists of a sequence of relations between node types. Formally:

\vspace{-0.05cm}
\begin{definition} 
A {\bf meta-path} or {\bf predefined-path} $P$ on a heterogeneous graph $G$ with object types $A_1, A_2, \dots$ and relations $R_1, R_2, \dots$: 

\begin{center}
\vspace{-0.2cm}
{ \footnotesize $P = A_1 \dots A_t = A_1 \xrightarrow{R_1} A_2 \xrightarrow{R_2} A_3 \xrightarrow{R_3}  \dots \xrightarrow{R_{t-1}} A_t$}
\vspace{-0.2cm}
\end{center}

 defines a composite
relation {\footnotesize $R = R_1 \circ R_2 \circ \dots \circ R_{s-1}$} between the source node type $A_1$ and the target node type $A_t$, where
$\circ$ is the composition operator on relations. 
\end{definition}

In our setting, the object types include users $U$, movies $M$, and features $F$, and there exist two types of relations: `rated-by' and `belongs-to', as well as their inverses. 

To capture user-feature preferences, we propose a random-walk-based score restricted over predefined paths  starting from $F$ and ending in $U$. We view the proximity of a user to a feature through such a path as an indication of the user's possible `preference' towards this feature. For example, a 2-step path 
via movies captures \textit{direct} user-feature preferences based solely on the movies they have rated. Unlike our approach, the original path similarity~\cite{SunHYYW11} is computed as a normalized count of paths between entities. For different types of preferences, we consider three predefined paths:
\vspace{-0.2cm}
\begin{itemize*}

\item \textbf{ 2-step Path}: The path $P = FMU$ in Figure~\ref{fig:twostep} finds the preferences of each user based on her  ratings, and, thus, does not exploit a collaborative setting. It computes accurate preferences for the features that appear in movies she has rated, but does not infer preferences for other features. 

\item \textbf{ User-based 4-step Path}: The path $P = FMUMU$ in Figure~\ref{fig:fourstep1} computes the user preferences based on the user's input, as well as that of other users who are \textit{similar} to her. The similarity between users is defined via the movies they have watched; two users who have rated the same set of movies with comparable scores are similar.

\item \textbf{ Feature-based 4-step Path}: The path $P = FMFMU$ in Figure~\ref{fig:fourstep2} finds the user preferences based on the user-movie ratings and \textit{similar} movies to the ones she rated. The similarity between movies is content- or feature-based. 
\end{itemize*}

One way to compute a user-feature score is through  the probability that a random walk starting at a given node, restricted to follow only paths of the above three forms, terminates at a given feature. Computing this probability corresponds to matrix multiplications involving the transition matrices induced by each bipartite graph (i.e., user-movies and movie-features). 
  \featmodel improves upon this approach in two ways, by considering `positive' and `negative' random walks, and incorporating edge weights, as described below.

\vspace{0.2cm}
\noindent\textbf{Positive and Negative Walks.} One issue with the above approach for preference inference is that the random walk model cannot handle `dislikes' or `negative' weights. Constructing the user-movies graph by creating an edge for each rating results in feature scores that only increase in magnitude despite the fact that a rating may be below average,  indicating a  `dislike'. However, this is not desired in our \method problem; we want the features that contribute to movies disliked by a user to be penalized by receiving smaller scores than features that contribute to her favorite movies. 

To handle this case, we introduce the concept of \positive and \negative walks. Since the movie ratings, which incorporate `dislike' information, appear in the bipartite user-movie graph, we focus only on that. To obtain the \positive graph from the original bipartite graph, we derive a graph that captures `like' relations~\cite{DernbachTKWDA16}, i.e.\ edges with weight (rating) above the corresponding average user rating. Similarly, we obtain the \negative graph by capturing only `dislike' relations. The process of generating the \positive and \negative graphs is illustrated in Figure~\ref{fig:toy_example_posneg}.

\begin{figure}[t!]
\vspace{-0.3cm}
        \centering
        \subfigure[Original graph]{\label{fig:ratingsGorig} \includegraphics[width=0.29\columnwidth]{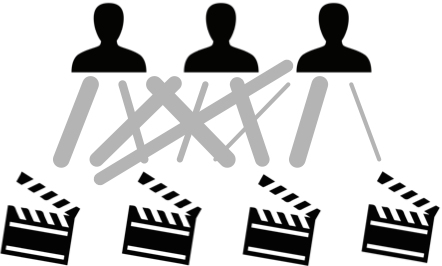}}
        ~
        \subfigure[{\scriptsize \positive} graph]{\label{fig:ratingsGpos} \includegraphics[width=0.29\columnwidth]{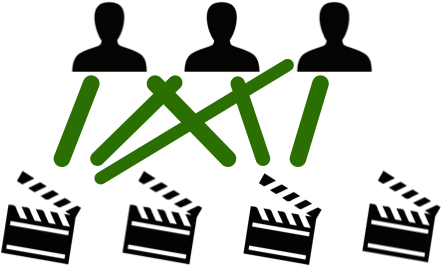}}
        ~
        \subfigure[{\scriptsize \negative} graph]{\label{fig:ratingsGneg} \includegraphics[width=0.29\columnwidth]{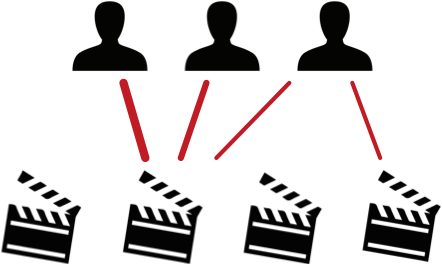}}
        
        \subfigure[Numeric Example]{\label{fig:numericExample} \includegraphics[width=0.95\columnwidth]{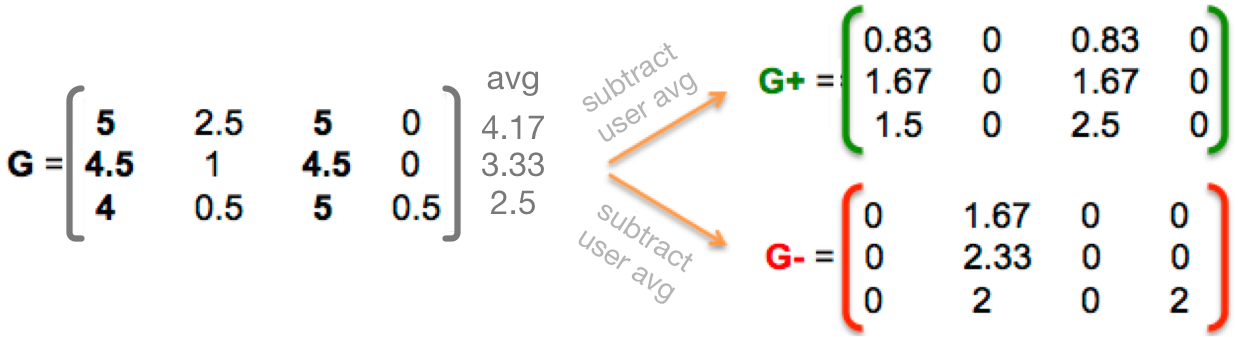}}
        \vspace{-0.3cm}
        \caption{Toy example of \positive and  \negative user ratings graphs for our proposed method \randomWalk.}\label{fig:toy_example_posneg}
\vspace{-0.3cm}
\end{figure}

\vspace{0.2cm}
\noindent\textbf{Edge Weights.} To better leverage the user preferences, we assign weights to edges between users and movies on both the \positive and \negative graphs, so that the walker performs a walk over a weighted graph.  
Since user ratings are often biased~\cite{koren2009matrix} (e.g., some users give high scores objectively, even if they did not enjoy the movie), we do not directly use the ratings as edge weights, but we center user $i$'s rating for movie $j$ through $r_{ij} - \bar{r}_i$, where $\bar{r}_i$ is the user's average rating. 
That is, we adjust the ratings according to each user's average movie rating and refer to it as a `Centered Rating' (CR). Second, to emphasize big deviations from the average user rating, we use a non-linear reweighing schema:

{
\begin{equation}
q_{ij} = 2^{\delta*(\text{CR})} = 2^{\delta \cdot |r_{ij} - \bar{r}_i|}
\label{eq:reweighing}
\end{equation}
}
which puts more weight on very low and very high ratings, where low and high are subjective for each user. We denote the resulting $u\times m$ weighted adjacency matrix by $\mathbf{Q}$, and consider random walks over the corresponding weighted graph. We study the effect of the reweighing schema and its non-linearity in Section~\ref{sec:experiments}.

\vspace{0.2cm}
\noindent\textbf{\featmodel Score Computation.} Putting all this together, \featmodel proceeds as follows. It computes  the probability that a random walk starting at a user reaches a feature, under the following assumptions:  the walk is (a) constrained to occur only over paths of the three proposed predefined types, and (b) occurs on either the \positive or the \negative graph, with edges appropriately weighted through Equation~\eqref{eq:reweighing}. Let $\vec{w_p}^{+}$ and $\vec{w_p}^{-}$ be the probability vectors over features, for path type $p$ and  user  $i\in \mathcal{U}$. Then,  we define $\vec{w_p} = \vec{w_p}^{+} - \vec{w_p}^{-}$ as the weighted user-feature preference vector over path type $p$. 
It is clear that, as desired, \randomWalk penalizes features in movies that are disliked by a user, by subtracting some value from their preference score. 
We define the final weighted feature vector of \randomWalk for $i$, which represents $i$'s preference over features in $\featureSet$ as a linear combination of the weighted user-feature vectors over the three predefined path types: 
\begin{equation}
\vec{w} = \alpha \, \vec{w}_{\text{2step}} + \beta \, \vec{w}_{\text{4step-usr}} + \gamma \, \vec{w}_{\text{4step-feat}} 
\end{equation}

where $\vec{w} = (\vec{w}^+ - \vec{w}^-)$. In matrix form, the feature preferences for all users are: 
{
\begin{equation}
 \mathbf{W} = \alpha \, \mathbf{W}_{\text{2step}} + \beta \, \mathbf{W}_{\text{4step-usr}} + \gamma \, \mathbf{W}_{\text{4step-feat}}, 
\label{eq:paths}
\end{equation}
}
\noindent where $\alpha$,$\beta$,$\gamma\in\mathbb{R^+}$ are the combination parameters satisfying the equation  $\alpha$+$\beta$+$\gamma$=$1$, and $\mathbf{W}$ is a  $u\times f$  matrix.

\begin{table*}[t!]
  \centering
  \caption{Dependencies between movie factors do not seem to correlate with movie success.}
{\footnotesize
\begin{tabular}{l|rrrrr|rr}
    \toprule
       & \multicolumn{5}{c|}{\textbf{Lasso Regression (5-fold C.V.)}} & \multicolumn{2}{c}{\textbf{Linear Regression  (5-fold C.V.)}} \\
    \multicolumn{1}{l|}{\textbf{Features}} & \multicolumn{1}{c}{\textbf{MSE (std)}} & \multicolumn{1}{c}{\textbf{non-0 coeff. (total)}} & \multicolumn{1}{c}{\textbf{Singletons}} & \multicolumn{1}{c}{\textbf{Pairs}} & \multicolumn{1}{c|}{\textbf{Triplets}} & \multicolumn{1}{c}{\textbf{MSE (std)}} & \multicolumn{1}{c}{\textbf{non-0 coeff.}}\\ \midrule
    singletons & 1.0395 (0.0533) & 211 (621) & 211 & n/a & n/a & 1.5448 (0.0467) & 620\\
    singletons + pairs & 1.0303 (0.0328) & 207 (969) & 161 & 46 & n/a & 3.025 (0.3498) & 899\\
    only pairs & 1.1103 (0.0327) & 141 (348) & n/a & 141 & n/a & 4.0232 (17.3369) & 100\\
    singletons + pairs + triplets & 1.0452 (0.0526) & 235 (1104) & 179 & 45 & 11  &  2.9254 (0.8767) & 951\\
    \bottomrule
    \end{tabular}
    }
  \label{tab:lasso_linear_reg}
  \vspace{-0.3cm}
\end{table*}

\subsection{Step 2. Designing the movie}
\label{seq:step2}

The second constituent problem of \method, Problem~\ref{prob:step2}, tackles the actual selection of features, $\mathcal{S}$, for a new movie so that the number of users who will likely enjoy it is maximized. We propose to formulate Problem~\ref{prob:step2} as an optimization problem with  cardinality constraints (and, if available, budget constraints). Following the literature on node-specific threshold models for influence maximization~\cite{KempeKT03,ChenYS09}, we consider a user $i$ converted with respect to (i.e., `likes') a movie consisting of features $\mathcal{S}$ if $\sum_{k \in S} w_{ik} > \thres_i$, where $w_{ik}$ is user $i$'s preference score for feature $k$ and $\tau_i$ is the user-specific threshold. Treating $\tau_i$ as a random variable, we define the user conversion function, $G(\selectedSet)$, as the expected number of  users in the target set $\userSet'$  that are converted:  

{\small 
\begin{equation}
G(\selectedSet) = \sum_{i \in \userSet'}P[\sum_{k \in \selectedSet} w_{ik} > \thres_i] = \sum_{i \in \userSet'}F[\wvec_{i}\xvec^T]
\label{eq:conversion}
\end{equation}
}
where $\xvec\in \{0,1\}^{f}$ is the characteristic vector of set $\mathcal{S}$ (i.e., $\xvec_k=1$ iff $k\in \mathcal{S}$). Based on these assumptions, we frame the \method task as a general  optimization problem:

\vspace{-0.3cm}
{
\begin{align}
\label{prob:opt}
& \text{Maximize}
& & G(\selectedSet) = G(\xvec)= \textstyle{\sum_{i \in \userSet'}} \; F[\wvec_{i}\xvec^T]  \\ 
& \text{subject to}
&& \textstyle{\sum_{k\in \mathcal{F}_\ell}}\; b_k \; x_k \leq B_\ell, \quad\text{for each }\ell\in\mathcal{T},\\ 
&&& \xvec\in \{0,1\}^f. \nonumber
\end{align}
}
\vspace{-0.4cm}

In the special case where $b_k=1$ for all $k\in \mathcal{F}_\ell$, the above problem effectively amounts to maximizing $G(S)$ subject to a \emph{cardinality constraint} $|\mathcal{S}\cap \mathcal{F}_\ell|\leq B_\ell$.  Such constraints also make sense as, e.g., no more than one director is needed per movie. If the cost $c_k$ per feature is available, then we can set $b_k=c_k$, and, thus, capture budget constraints. 
We note that the optimization problem has a linear constraint, and the objective depends on the inner product of scores with $\xvec$; as such, it does not model dependencies between features. Next, based on evidence from the real-world data, 
we explain why we do not model such dependencies.

\vspace{0.1cm}
\noindent\textbf{Why not model feature dependencies?} 
We followed a data-driven approach to evaluate whether dependencies between movie features correlate with movie ratings (e.g., two actors should always play together because then they lead to successful movies, while independently they do not). 
Using the IMDB dataset described in Section~\ref{sec:data}, 
we performed: (1) frequent item mining (a priori algorithm) to find frequent $k$-itemsets that have appeared in at least ten movies; and then (2) linear and lasso regression with 5-fold cross validation to select the $k$-itemsets (singletons, pairs or triplets of features) that are most predictive of the movie rating. 

Frequent item mining showed that only 1\% of the features are frequent (with support $\geq0.5\%$), half of which contribute to frequent pairs, and 20\% contribute to frequent triplets. The results of linear and lasso regression are summarized in Table~\ref{tab:lasso_linear_reg}. In general, regression on the dataset with frequent singleton features is better or comparable (in terms of error and model compactness) to regression on datasets that capture feature dependencies. Specifically, the MSE (Mean Squared Error) of linear regression  is minimum when the input dataset has only singleton features. The MSE of lasso regression becomes slightly smaller when we consider frequent pairs of features rather than only frequent single features (1.04 vs.\ 1.03). However, the majority of selected features are singletons, and the number of selected features are almost the same. Therefore, the feature dependencies are very few and do not affect the predictability of  ratings significantly.

\vspace{0.2cm}
\noindent\textbf{Uniform Threshold Model.} We propose to use a linear threshold model where each user $i$ picks her conversion threshold $\tau_i$ uniformly at random~\cite{KempeKT03}, i.e., we assume no background information about the user conversion thresholds. Since the \randomWalk user-feature preferences computed  are in $[-1,1]$, we set the thresholds to follow the uniform distribution in the same interval, i.e. $\thres_i \sim \mathcal{U}[-1,1]$. Under these assumptions, the user conversion function~\eqref{eq:conversion} becomes 

{
\begin{equation*}
f(S) = \frac{1}{2} \onevector^T \wMat \onevector + \frac{|\userSet|}{2} 
\end{equation*}
} 
\vspace{-0.2cm}

\begin{proof}
Starting from Equation~\eqref{eq:conversion} and applying our assumptions, we obtain:
{\footnotesize
\begin{align*}
f(S) &= \textstyle{\sum_{i \in \userSet'}}P[\textstyle{\sum_{k \in \selectedSet}} w_{ik} > \thres_i]
  \stackrel{ \thres_i \sim \mathcal{U}[-1,1]}{=} \\
&=\textstyle{\sum_{i \in \userSet'}} \frac{1}{2}(\textstyle{\sum_{k \in \selectedSet}} w_{ik}+1) = \\
&= \frac{1}{2} \textstyle{\sum_{k \in \selectedSet}}\textstyle{\sum_{i \in \userSet'} w_{ik}}+\frac{|\userSet|}{2} 
\stackrel{ w_k=\textstyle{\sum_{i \in \userSet'}} w_{ik} }{=\joinrel=\joinrel=} \\
& = \frac{1}{2}  \textstyle{\sum_{k \in \selectedSet}} w_{k} + \frac{|\userSet'|}{2} 
\end{align*}
}
Hence, under the uniform threshold model, we want to pick the movie features that maximize $\sum_{k \in \selectedSet} w_k $. 
\end{proof}

In this case, the optimization problem reduces to knapsack under several separable constraints, which, though NP-hard, can be solved with an FPTA scheme \cite{vazirani2013approximation}. In the unit-cost (i.e., cardinality) case, the problem is solvable in polynomial time:
sorting features $k\in \mathcal{F}$ in decreasing order of costs ${w_k}$, and picking the top $B_\ell$ features, is optimal. 

\vspace{0.2cm}
\noindent\textbf{Sigmoid Threshold Model. } The uniform threshold model has the advantage of resulting in an easy-to-solve optimization problem. To further leverage background information that might be available in the data, data-driven models may be considered for the user-specific thresholds. For instance, if the average rating score per user 
follows a sigmoid distribution and it is being used as the user-specific threshold, then $F$ in Equation~\eqref{eq:conversion} can be replaced with the CDF of the logistic distribution, $F_S$. 

This results in a sigmoidal programming problem~\cite{UdellB14sigmoids}, 
which is NP-hard even in its relaxed form with non-integral solution. 
By formulating the problem as the maximization of $\sum_{i \in \userSet}F_S[y_i]$, we introduce an additional set, $|\featureSet|$, of constraints $y_i = \wvec_{i}\xvec^T$,  which can be solved approximately by using the branch-and-bound based method in~\cite{UdellB14sigmoids}. The decimal solutions can then be converted to integral by using pipage rounding~\cite{AgeevS04}. However, the approximation error of the solution depends on the number of constraints, and 
the method 
may solve many convex optimization problems exponentially, which suggests that the method will be impractical for the \method problem with    thousands of constraints.

\subsection{Complexity of \featmodelT}
\label{sec:scalability_theory}
\vspace{0.05cm}
\noindent \textbf{Na\"ive approach.} Performing the walks over the 2-step and 4-step predefined paths on the positive and negative graphs is computationally expensive and is dominated by sparse matrix and sparse-dense matrix multiplications 
(e.g., $\tilde{\matQ} \tilde{\matF} $ for $P = FMU$, or  $\tilde{\matQ}\tilde{\matQ}^T\tilde{\matQ}\tilde{\matF}$ for $P = FMUMU$, 
where $\tilde{\matQ}, \tilde{\matQ^T}, \tilde{\matF} $ are $u \times m$,  $m \times u$, and $m\times f$ transition matrices). If we assume na\"ive matrix multiplications, the complexity of performing the walks on the 2-step, user-based 4-step, and feature-based 4-step paths would be: $\Theta(fmu)$, $\Theta(fmu + 2u^2m)$ and $\Theta(3fmu)$. We note that these computations can be trivially parallelized, so the complexity 
corresponds to the maximum of the three. Moreover, the output is a dense $u \times f$ matrix $\mathbf{W}$ that captures the user-feature preferences. 
Although the bulk of the computation can happen offline, we can support quick query processing by just summing up the rows in matrix $\mathbf{W}$ that correspond to the target users $\userSet'$ and performing linear feature selection in 
$O(u' f)$ after sorting the scores in decreasing order in $O(f \xspace \log f)$. 
Next, we show that we can significantly speed up the computation and reduce the storage requirements.  

\vspace{0.2cm}
\noindent\textbf{Fast approach.} Noticing that the uniform threshold model requires only the sum of the target users' feature preferences (and not the whole $\mathbf{W}$ matrix)  helps us to significantly speed up the computations and reduce the storage requirements. By using an indicator vector $\vec{x}$ for the users in $\userSet'$, all the walks can be re-designed (backwards) as fast sparse matrix-vector multiplications---for example, 
{ $\tilde\matF^T( \tilde\matQ^T( \tilde\matQ (\tilde\matQ^T \vec{x})))$}
in { $\Theta((3 \;\textrm{nnz}(\tilde\matQ)$ $+ \textrm{nnz}(\tilde\matF)) u')$}, where nnz() is the support size (i.e., number of non-zeros) of the corresponding matrix. The transpose of the resulting score matrix is $\mathbf{W}$. The intermediate and final computations generate and store vectors of variable lengths. The feature selection can be done in $O(f \xspace \log f + f)$. 
As we show  in Section~\ref{sec:scalability_practice}, the fast approach is very scalable and up to $100\times$ faster than the na\"ive approach.

\section{Data}
\label{sec:data}
We compiled our data from two different and publicly available sources: (i) the Flixster website~\cite{ZafaraniL09}, a social movie site allowing users to watch and rate movies, from which we obtained user-movie ratings, and (ii) the MGO website~\cite{mgo} which has information about movie features. The Flixster and MGO datasets consist of 66,726 and  83,017 movies, respectively. In addition to the user ratings, the Flixster dataset provides some user demographics, including their gender, age, and location (as free text).

To merge the two datasets, first we dropped the movie remakes. 
In order to be able to infer the user-feature or user-movie preferences (Problem 1) using 5-fold cross validation, we iteratively filtered the data so that:
every movie has been rated by at least 20 users and has at least 2 features; every user has rated at least 20 movies; 
every feature has participated in at least two movies.
The resulting user-movie-feature tripartite graph has $66,407$ nodes and $4,567,253$ edges 
(Table~\ref{tab:filtered}). 
The movie features include: $25,721$ actors,  $1,322$ directors, $4,305$ producers, $654$ studios, and $27$ genres. 
The user ratings are between 1 and 5, with 0.5 increments. 

For our dependency analysis we used an IMDB dataset consisting of $1,893$ movies, their features ( $947$ directors, $5,231$ producers, $1,209$ studios, $51,500$ actors, $27$ genres, and $4$ seasons), and average movie ratings.

\begin{table}[bth!]
\vspace{-0.4cm}
  \centering
  \caption{Statistics of our movie dataset.}
  \resizebox{\columnwidth}{!}{
    \begin{tabular}{lr||lr}
    \toprule
    \textbf{Movies}	  & 5,881  & \textbf{User-Movie Ratings} & 4,435,359 \\
    \textbf{Users}    & 28,482  & \textbf{Movie-Feature Memberships} & 131,894\\
    \textbf{Features} & 32,029 & & \\ 
    \bottomrule
    \end{tabular}
    }
  \label{tab:filtered}
\end{table}
\vspace{-0.2cm}

\section{Experiments} 
\label{sec:experiments}

In this section, we give the experimental analysis of \randomWalk, and evaluate its performance for designing new movies for specific audiences. We seek to answer four main questions:
\vspace{-0.1cm}
\begin{itemize*}
\item[\textbf{Q1.}] How does \featmodel compare to the baseline approaches? 
\item[\textbf{Q2.}] Is \featmodel robust with respect to its parameters? 
\item[\textbf{Q3.}] Is \randomWalk scalable? 
\item[\textbf{Q4.}] How successful are the movies we  design? 
\end{itemize*}
The first three questions are related to the inference of user-feature-score preferences (Problem 2), and the last question refers to the optimization problem 
(Problem 3).

\subsection{Analysis of Step 2: Inferring feature scores}
\subsubsection{Q1: \featmodelT vs.\ baseline methods}

To compare the predictive power of \featmodel against baseline methods, we cast Problem~\ref{prob:step1} as a binary classification problem where the goal is to predict whether a given user will like or dislike a new movie. More formally, recall that 
we consider 
two classes~\cite{DernbachTKWDA16}: 
User $i$ likes (dislikes) movie $j$ if she rated it higher (lower) than her average user score, i.e., $C_{ij}$=$1$ if $r_{ij} \geq \bar{r}_i$ ($C_{ij}$=$0$ if $r_{ij} < \bar{r}_i$). 
Each observation corresponds to a user-movie rating, where the independent variables are the movie features (binary), and the dependent variable 
is the user's preference (like or dislike). 
We note that the mapping of the user-ratings to binary values results in 55\% 
`liked' movies, and 
45\%
`disliked' movies.  
In all the cases that we describe below, we learn a per-user classifier on their movie ratings, and test it on new movies. 
To predict the user's preference for 
a new movie using \featmodel, we  multiply the user-based feature-preference vector $\mathbf{w}_i$ with the new binary movie vector, and compute 
her movie score. 

This problem can be solved by several traditional methods, among which we consider:

\vspace{0.2cm}
\noindent{\bf Baseline 1.1: Naive Bayes.} NB can predict whether or not a user will like a new movie, i.e., it cannot infer the user-feature preferences (Problem~\ref{prob:step1}). NB makes the assumption that the features act independently. 
 To avoid the zero-probabilities issue, we use the Laplace correction.

\vspace{0.2cm}
\noindent{\bf Baseline 1.2: Logistic Regression.} 
LR is a commonly-used generalized linear model for prediction. 
Similarly to NB, LR also assumes feature independence.  LR also generates a weight vector of the features' contribution in the prediction. 
Since the number of features is greater than the number of samples (Table~\ref{tab:filtered}), we used L1 regularization to ensure sparsity of the produced weight vector. To find the best value of the regularizer parameter, we performed a cross-validated grid-search over the values \{0.5, 1, 5, 10\}.

\vspace{0.2cm}
\noindent{\bf Baseline 1.3: Random Forests.} 
RF is an ensemble method for classification and regression, which is robust to the inclusion of irrelevant features and overfitting. 
 For each user, the RF classifier generates N random subsets on the training data in terms of features, and constructs a separate decision tree for each subset. To predict a new user-movie preference, the movie's feature vector (new sample) runs through all the generated user-specific decision trees, and the mode of the classes is assigned as the predicted class. 
A drawback of this approach is its runtime, since it requires generating many decision trees per user. In our experiments, for the maximum depth of the trees, we performed grid-search over the values \{25, 100, 500, 1000\}, and generated 10 random subsets of the data. Overall, RF was very slow despite the small number of grid-search values.

\vspace{0.2cm}
\noindent{\bf Baseline 1.4: Matrix Factorization.} 
MF~\cite{koren2009matrix,bennett2007netflix} with user and item biases predicts the rating of user $i$ for item $j$:

\vspace{-0.2cm}

{\small
$$\hat{r}_{ij}= \left\langle u_i, v_j\right\rangle + \mu + b_i + b_j$$
\vspace{-0.4cm}
}

\noindent where $u_i$ and $v_j$ are $d$-length vectors (for some small $d$)  
that capture the user and item latent vectors, respectively,
$\mu$ is the global average rating, $b_i$ is the user bias, and $b_j$ is the item bias.
The main advantage over the other approaches is
that we learn a model for each user using a small dense matrix, where the number
of rated movies is commonly larger than the number of features. However, 
MF operates on a latent feature space,
thus using it to design a new movie 
is not straightforward. 
We performed MF  over the 
ratings matrix $\matR$ to compute user and feature
vectors with a dimension of $d=10$. These were computed using 20 iterations of stochastic gradient descent, and $\ell_2$
regularization parameters selected through cross validation. 
The predicted ratings of user/movie pairs were used in the AUC computations.

\vspace{0.2cm}
\noindent{\bf Baseline 1.5: Content-based Matrix Factorization.} 
We further extended basic MF by incorporating content-dependent biases: beyond user and movie biases, we include additional bias in our model, one for  every item feature. These can be thought of as explicit, rather than latent, features of a movie. We again compute parameters for $\ell_2$ regularization penalties through cross validation.

\vspace{0.2cm}
\noindent{\bf Baseline 1.6: Heterogeneous Entity Recommendation.} 
The method in~\cite{YuRSGSKNH14} finds the top-$k$ movies to recommend to a user by performing non-negative matrix factorization on a set of user-movie preference diffusion matrices learned via meta-path similarity~\cite{SunHYYW11} and user clustering.  
We note that this approach cannot be used for \textit{feature} preference inference since the model 
applies only for entities that are directly linked to and rated by users in the heterogeneous graph (e.g., movies). Thus, in the binary classification task, we use the inferred user-\textit{movie} scores which are directly learned over the metapaths that correspond to our proposed method's predefined paths. We set $k=20$ for the low-rank approximations and pick the method's parameters through  cross validation.

\vspace{0.2cm}
\subsubsection*{Results} 
 To evaluate the methods, we perform 5-fold cross validation, compute the AUC (Area Under the Curve) per user, and report its average and standard deviation over all users in Table~\ref{tab:crossVal}.

\vspace{-0.2cm}
\begin{observation}
\featmodel outperforms all baseline approaches on user-feature preference inference. 
\end{observation}
\vspace{-0.2cm}

The main shortcomings of the baselines are: (i) Each per-user model for NB, LR, and RF leverages information specific to that user only---i.e., her movie ratings, and the corresponding movie features---, and thus suffers from the sparsity problem; (ii) MF falls in the collaborative space, but does not leverage information about the movie features, and operates on a latent feature space; content-based MF also partially relies on latent movie features, which again prohibits its use in movie design. The heterogeneous entity recommendation approach infers the movie preferences, but not the feature preferences, and, thus, cannot be used for movie design either. In contrast, our proposed method, \featmodel, solves the \method problem by exploiting the latent similarities between users (collaborative), movies, and features, and operates explicitly on the movie features, thus directly enabling us to use its output to design movies.

\begin{table}[t!]
\vspace{-0.2cm}
\centering
\caption{\randomWalk outperforms all the baselines. We report the prediction accuracy (avg AUC and its std in parentheses). $\alpha, \beta, \gamma$ are the path combination parameters, and $\delta$ is the rating-reweighing parameter. }
\label{tab:crossVal}
\resizebox{\columnwidth}{!}{
\begin{tabular}{ l r }
\toprule
\textbf{Method} & \textbf{Avg AUC over 5 folds}\\ 
\midrule
\bf{Naive Bayes} & 0.5340 (0.0002)  \\
\bf{Logistic Regression} & 0.6621 (0.1369)  \\
\bf{Random Forests} & 0.6418	 (0.1390)  \\
\hline
\bf{Matrix Factorization} & 0.6396 (0.0825) \\
\bf{Content-based Matrix Factorization} & 0.7043 (0.1420)\\
\bf{Heterogeneous Entity Recommendation} & 0.5998 (0.0112) \\
\hline
\bf{\featmodel} & \\
$(\alpha,\beta,\gamma,\delta)=(0.7, 0.1, 0.2, 1)$ &  0.9146	(0.0043)\\
$(\alpha,\beta,\gamma,\delta)=(0.5, 0.2, 0.3, 1)$ & 0.9143	(0.0043) \\
$(\alpha,\beta,\gamma,\delta)=(0.3, 0.2, 0.5, 1)$ &  0.9132	(0.0042) \\
$(\alpha,\beta,\gamma,\delta)=(0.5, 0.2, 0.3, 0.5)$ & \textbf{0.9171}	(0.0019)  \\
$(\alpha,\beta,\gamma,\delta)=(0.5, 0.2, 0.3, 1.5)$ &  0.9062	(0.0018) \\
\bottomrule
\end{tabular}
}
\vspace{-0.4cm}
\end{table}

\subsubsection{Q2: Robustness of \featmodelT}

\begin{figure*}[t!]
	\vspace{-0.2cm}
	\centering
	\subfigure[Robustness wrt reweighing schema\label{fig:exponent_analysis}]{\includegraphics[width=0.31\textwidth]{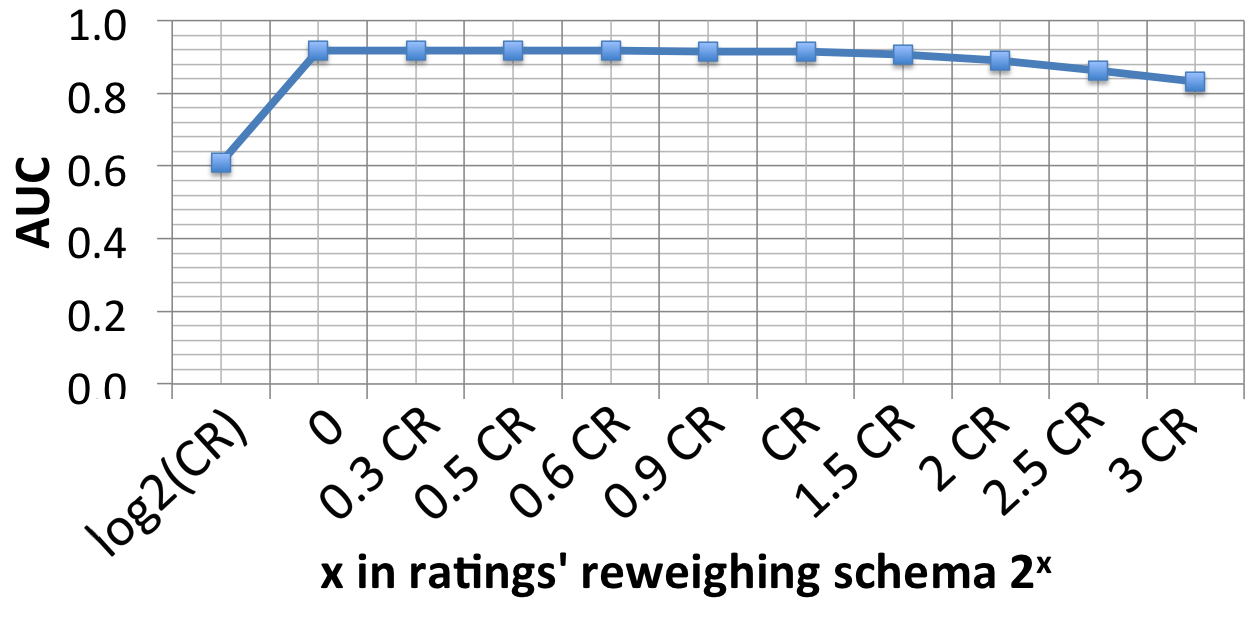}}
	~
	\subfigure[Robustness wrt combination params \label{fig:robustness-param}]{\includegraphics[width=0.31\textwidth]{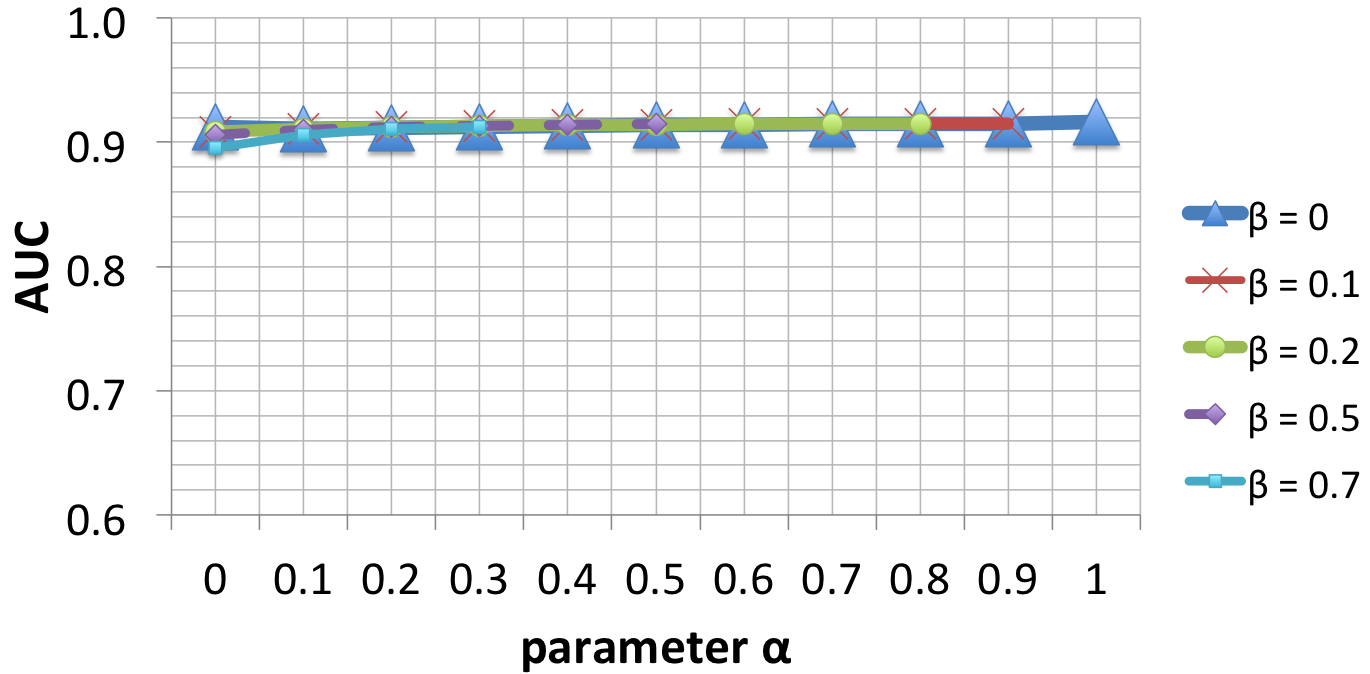}}
	~
	\subfigure[Scalability \label{fig:scalability}]{\includegraphics[width=0.31\textwidth]{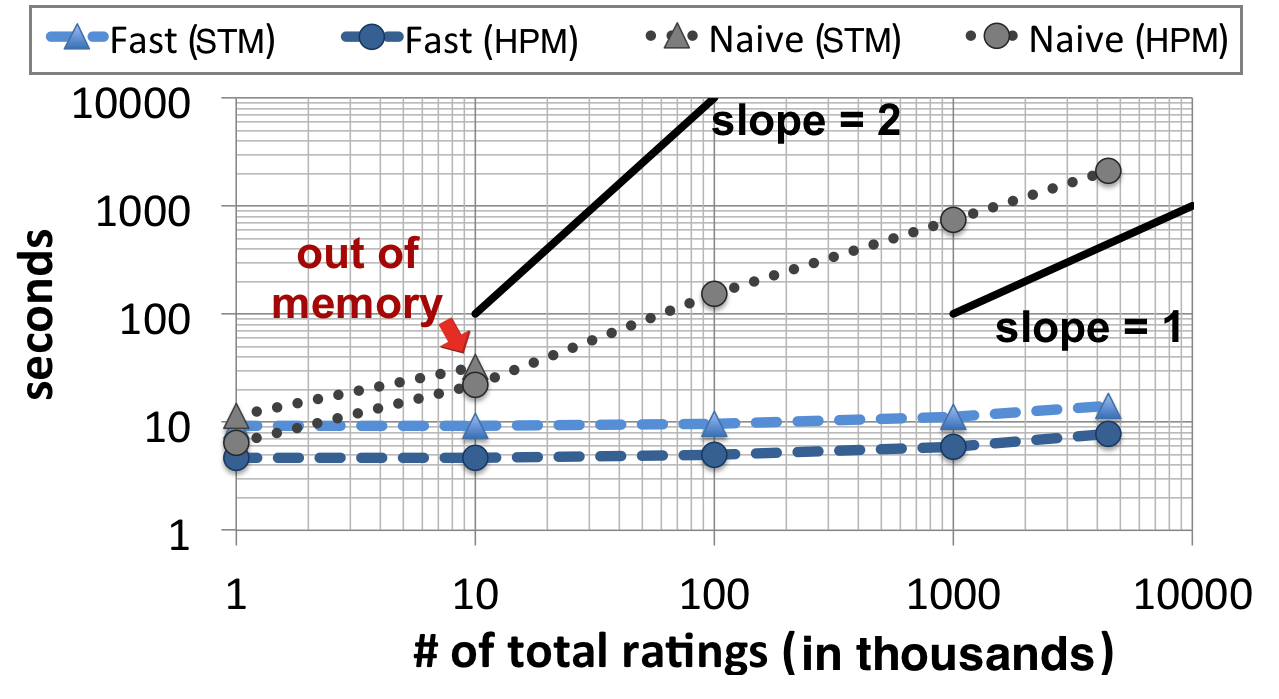}}
    \vspace{-0.2cm}
	\caption{\subref{fig:exponent_analysis} \featmodel is robust to moderate rescaling of the centered ratings---x-axis: exponent of our proposed reweighing schema in Eq.~\ref{eq:reweighing}; y-axis: avg AUC over 5-fold cross validation. \subref{fig:robustness-param} \featmodel is robust to the combination parameters of the  predefined paths---x-axis: values of parameter $\alpha$; y-axis: avg AUC. \subref{fig:scalability} \featmodel is scalable. The fast approach is faster and needs less space than the naive approach. }
	\vspace{-0.35cm}
\end{figure*}

\vspace{0.2cm}
\noindent\textbf{Robustness to the reweighing schema.}
To evaluate our reweighing schema in Eq.~\ref{eq:reweighing}, we learn a \featmodel classifier per user for different values of the parameter $\delta$. 
In Figure~\ref{fig:exponent_analysis}, we give the average AUC (5-fold cross validation) of \featmodel for using the centered ratings without exponentiation (simple reweighing), and our proposed approach with $\delta$ ranging from $0$ to $3$, and for combination parameters $\alpha = 0.5$, $\beta=0.2$ and $\gamma=0.3$. The performance is stable for $\delta=0-1.5$, which corresponds to moderate scaling of the user ratings.

\vspace{-0.2cm}
 \begin{observation}
 \featmodel is robust to moderate rescaling of the user-adapted movie ratings.
 \end{observation}
 \vspace{-0.2cm}

\vspace{0.2cm}
\noindent\textbf{Robustness to the combination parameters.} 
As described in Eq.~\eqref{eq:paths}, the proposed method, \featmodel, has three parameters which define the involvement of each predefined path. Next, we explore how 
these parameters affect the user preference inference. 
As before, we learn a \featmodel classifier per user for different values of $\{\alpha, \beta, \gamma\}$, and compute the average AUC 
and its standard deviation (y-axis) by performing 5-fold cross validation (Figure~\ref{fig:robustness-param}). In this experiment we set 
$\delta=1$. 
We vary the values of $\alpha$ and $\gamma$ from $0$ to $1$ with step $0.1$ ($\beta$ is uniquely determined since $\alpha + \beta + \gamma = 1$).

\vspace{-0.2cm}
 \begin{observation}
 \featmodel is quite robust to the `combination' parameters ($\alpha, \beta, \gamma$) of the predefined walks.
 \end{observation}
\vspace{-0.2cm}

\new{The lowest accuracy is obtained when the direct user preferences (2-step path) are ignored, and the highest accuracy when the user-based 4-step path has small participation, likely because this walk `blurs' the individual preferences by relying on `similar' users. In practice, these parameters can be set by performing cross validation on the data at hand.}

\subsubsection{Q3: Scalability}
\label{sec:scalability_practice}

We evaluate the scalability of our method with respect to the number of ratings, which corresponds to the number of non-zeros in the ratings matrix $\matR$. For our experiment, we vary the number of ratings from $1000$ to $4\,435\,359$, which is the total number of ratings in our dataset. For each number of ratings, we generate five matrices with randomly chosen ratings from the original matrix, we run \featmodel and report the average and standard deviation of the runtime. For comparison, we ran both the na\"{i}ve and fast approaches (we give their theoretical complexities in Section~\ref{sec:scalability_theory}). Figure~\ref{fig:scalability} shows the runtime of the two methods on: 
\vspace{-0.1cm}
\begin{itemize*}
\item[(a)] a standard machine (STM): AMD Opteron Processor 854 @ 2.80GHz, 2 cores, 32GB RAM;
\item[(b)] a high-performance machine (HPM): AMD Opteron Processor 6282 SE @ 2.60GHz, 16 cores, 264GB RAM. 
\end{itemize*}
 We note that the na\"ive approach runs out of memory for more than $10\,000$ ratings on the standard machine, while our fast approach is highly scalable and has comparable runtime on both machines.

\vspace{-0.2cm}
\begin{observation}
The fast approach (sparse matrix-vector multiplications) is up to $100\times$ faster than the na\"{i}ve approach, and requires less memory.
\end{observation}
\vspace{-0.2cm}

\subsection{Analysis of Step 3: Designing the Movie}

\subsubsection{Q4: Quantitative Evaluation}

To evaluate the quality of our movie designs quantitatively, we compare it with two intuitive baselines:
\vspace{-0.1cm}
\begin{itemize*}
\item \textbf{Baseline 2.1: }\popular. Using the same capacity constraints as in the optimization problem (Eq.~\ref{prob:opt}), we design a movie by choosing the most popular features for the target audience $\userSet'$ 
(the ratings are inherited from the movies to which they belong). The $1\times f$ popularity vector is given by: $\vec{p} = \vec{v} \cdot \matF$, where $\vec{v}$ is the vector of total ratings per movie, with elements $v_j = \sum_{i=1}^{u'}\mathbb{1}_{r_{ij}>0}$, $\mathbb{1}$ is an indicator function, and $\matF$ is the binary movie-feature membership matrix.

\item \textbf{Baseline 2.2:} \topF. Applying the same capacity constraints, we design a movie by selecting the most highly rated features for the target users  $\userSet'$. We assume that the features inherit their movies' ratings, so the vector of feature ratings can be computed as: $\vec{t} = \vec{\bar{r}_j} \cdot \widetilde{\matF}$, where $\vec{\bar{r}_j}$ is the vector with the average movie ratings over all users in $\userSet'$, and $\widetilde{\matF}$ is the column-normalized movie-feature membership matrix.  
\end{itemize*}

We introduce two quantitative measures to compare the \featmodel movie designs to the \popular and \topF designs. 
\vspace{-0.1cm} 
\begin{itemize*}
\item \textbf{\knn}: Given a movie $j^*$, 
we find the set $\mathcal{N}$ of its $k$ nearest neighbors (via cosine similarity) and infer its score 
as $\hat{r}_\vec{j^*} = \frac{1}{|\mathcal{N}|}\sum_{j \in \mathcal{N}} \bar{r}_j$, where  $\vec{\bar{r}}$ is the vector with the average movie ratings over all users in $\userSet'$. We set  $k=20$.

\item \textbf{\wknn}: This is the weighted version of \knn, where the movie scores are weighted by their cosine similarity to the given movie $j^*$, i.e., $\hat{r}_{j^*} = \alpha \sum_{j \in \mathcal{N}} sim(j^*, j) \cdot \bar{r}_{j}$, where $\alpha = 1 / \sum_{j \in \mathcal{N}} sim(j^*, j)$ is a normalization factor.

\end{itemize*}

\vspace{0.2cm}
\noindent \textbf{Results.} We applied all methods to the movie dataset (Section~\ref{sec:data}) in order to design movies (with capacity constraints, e.g., six actors, two directors, two genres) for different target audiences: (i) all women; (ii) all men; (iii) 13- to 15- year-old children; (iv) 13- to 18- year-old children; (v) men 18-25 years old; (vi) women 18-25 years old. 

\vspace{-0.2cm}
\begin{observation}
As shown in Figures~\ref{fig:evaluationmethod1} and~\subref{fig:evaluationmethod2}, \featmodel outperforms the competing baselines  \popular and \topF designs and obtain higher \knn and \wknn scores.
\end{observation}
\vspace{-0.2cm}

\begin{figure}[t!]
\centering
\subfigure[\knn scores per target audience.]{\includegraphics[width=0.95\columnwidth]{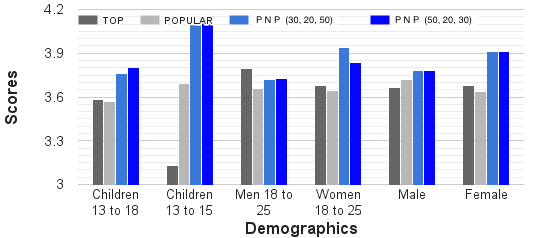}\label{fig:evaluationmethod1}}
\subfigure[\wknn scores per target audience.]{\includegraphics[width=0.95\columnwidth]{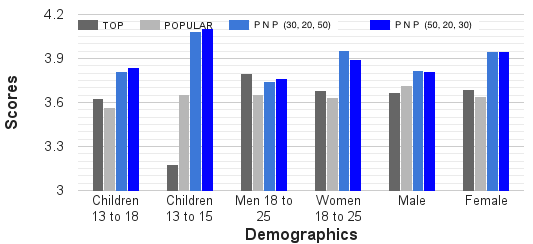}\label{fig:evaluationmethod2}}
\vspace{-0.2cm}
\caption{\featmodel outperforms the \textsc{TOP} and \textsc{POPULAR} movie designs. }
\vspace{-0.3cm}
\end{figure}

\subsubsection{Anecdotes}
To highlight how the movie design changes with the age of the target audience, we present the movies designed for men across
different age groups. 
For teenage boys (13-18years), the movie designed has the following composition. Top genres are Martial Arts and Action; top actors chosen are Will Smith (according to IMDB, best known for the Men In Black series) and Adam Sandler (known for Grown Ups); directors are Clint Eastwood (known for Million Dollar Baby) and Spike Lee (known for Malcolm X); producers are Mark Burg (known for Two and a Half Men) and Jack Giarraputo (known for 50 First Dates); studios are Revolution and Warner Brothers.
Interestingly, the top chosen genre for a movie for 19-25 year-old men is Romance, with Ned Bellamy (best known for Being John Malkovich) 
and Alan Rickman (known for the Harry Potter series) as the leading actors, and Ron Howard as the director (known for Cinderella Man). As another example, the movie designed for men of all ages has Martial Arts (the genre for movies like Rush Hour 2) as top-genre, Gary Oldman (known for the Batman series) and Matt Damon (Bourne series) as lead actors, and Steven Spielberg as lead director. 
Our model leverages the distinct patterns in the movie interests of different demographics to design movies that specifically
cater to those audiences. 
That said, movie making is a highly creative process, in which there are many elements that need to come together. Taking such a data-driven approach to designing movies is a step towards the world of targeted and personalized movies.

\section{Related Work}
\label{sec:related}
Next we review three main related areas: (i) recommendation systems, (ii) analysis of heterogeneous graphs, and (iii) influence maximization. 

\textbf{Recommendation systems.} Our work  is related to recommendation methods, and specifically group recommendation, which seeks to recommend \textit{existing} items that are likely to match the tastes of multiple users at the same time \cite{OConnorCKR01,JamesonS07}. The area of recommendation systems  is very active with numerous algorithms that leverage the user preferences and similarities~\cite{HuKD08,BreeseHK98}, or item similarities \cite{SarwarKKR01,LindenS03} in order to provide personalized recommendations to the users. The user-based CF approaches suffer from the sparsity problem, especially for new users (cold-start problem), which leads to very 
low prediction quality and scalability. To overcome these problems, \cite{MaYLK08} leverages the social interactions and connections between users, \cite{YuanCZ11} proposes a unified approach that combines CF with friendships and memberships,  
and content-based methods use side information, such as demographics, and item features (e.g., stylistic visual attributes) ~\cite{MelvilleMN02,Ronen2013RECSYS,deldjoo2016content}. 
A popular, scalable and accurate approach is matrix factorization \cite{koren2009matrix},
which relies on latent factors for both users and items. 
Privacy is always a major concern in online systems including recommendation systems, which has led to privacy-preserving systems~\cite{McSherryM09,BhagatWIT14}. 
Despite the similarities with group recommendation, our objective 
is to design a \textit{new} movie (or, generally, a new product), so that the expected number of endorsers in the intended audience  is maximized.

\vspace{0.1cm}
\textbf{Heterogeneous graphs.} Heterogeneous networks have become very popular in the recent years, and efforts have focused on adapting and extending approaches 
intended for homogeneous graphs, such as similarity search \cite{SunHYYW11} and random walk with restarts~\cite{BalminHP04,NieZWM05}. 
Sun et al.~\cite{SunHYYW11} introduced the idea of meta-paths for similarity search. In our work, we use meta-paths or predefined paths to perform `random' walks, with the ultimate goal of inferring user-feature preferences for the \method problem.   
The heterogeneous entity recommendation approach~\cite{YuRSGSKNH14} addresses the problem of top-$k$ entity recommendation by performing matrix factorization on user-movie preference matrices obtained by employing the definition of meta-path similarity~\cite{SunHYYW11} (normalized path counts). 
The goal of~\cite{YuRSGSKNH14} is to rank \textit{existing} items by user interest, assuming that all the items that are rated by a user are of interest to her (although she might have disliked some of the items). This is a weaker problem than the one that our work and typical recommender systems tackle. Moreover, \cite{YuRSGSKNH14} cannot directly infer the user-feature preferences, making it unsuitable for the \method problem.   
Unlike these works, we infer the user preferences with respect to item \textit{features} by defining random-walk scores over meta-paths and effectively incorporate likes and dislikes, which leads to high accuracy.

\vspace{0.1cm}
\textbf{Influence Maximization.}  Influence maximization is a fundamental underlying problem in viral marketing~\cite{ChenWW10}, the early adoption of products~\cite{Singer12},  targeted advertising, and more. The goal is to identify which users to target in order to maximize the adoption of an existing product, or service. 
One of the most influential papers in the area by 
Kempe, Tardos, and Kleinberg~\cite{KempeKT03} introduced the independent cascades and linear threshold models for user conversion; more efficient models have been proposed since then~\cite{ChenYS09}. In our work,  we use a variation of the linear threshold model for the \method problem.

\section{Conclusions}
\label{sec:conclusions}
This paper introduces the \textsc{Movie}-\textsc{Design} problem  for specific target audiences by leveraging user-movie preferences and movie content. 
 The \method problem that we have proposed is complementary to recommendation systems. In addition to contributing a novel formulation of the \method problem as an optimization problem, we have introduced a new random walk-based algorithm, \randomWalk on a \positive and \negative graph, which 
 efficiently handles dislikes in the user preferences. We have 
  shown that \randomWalk is superior to baseline methods in terms of movie preference predictions. Finally, we have applied our method on large, real-world datasets to generate movies for specific audiences, and introduced ways for the qualitative and quantitative evaluation of the designs. Although we have tailored our approach to the design of movies, it can be generalized to other products as long as reviews and product features can be identified. 

Future work includes extending our method with other elements of the creative movie process, notably the plot of the movie, the screenplay, and music. There are additional signals that can be incorporated, such as, the credits order for the actors in the movie as an indicator of their contribution and weights on a movie's genre, e.g.,\ for a movie that is mostly drama yet has an element of romance, we can weigh those two genres unequally.

{
\bibliographystyle{plain}
\bibliography{BIB/abbrev,BIB/danaiRef}
}

\end{document}